\documentclass[12pt]{article}
\usepackage{graphicx}
\usepackage{epstopdf}
\usepackage{amsthm}
\usepackage{amsmath}
\usepackage{amssymb}
\usepackage{amstext}
\usepackage{amsfonts}
\usepackage{enumerate}
\usepackage[authoryear,nonamebreak,bibstyle]{natbib}
\usepackage{footnote}
\makesavenoteenv{tabular}
\makesavenoteenv{table}
\usepackage[onehalfspacing]{setspace}
\usepackage[shortlabels]{enumitem}
\usepackage{multirow}
\usepackage[title]{appendix}
\usepackage{xcolor}
\usepackage{cancel}
\usepackage{hyperref}
\usepackage[normalem]{ulem}
\usepackage{tikz}
\usetikzlibrary{shapes,arrows}
\usepackage{tikz-cd}

\usepackage[numbib]{tocbibind}

\begin{document}
\baselineskip=.22in\parindent=30pt

\newtheorem{tm}{Theorem}
\newtheorem{dfn}{Definition}
\newtheorem{lma}{Lemma}
\newtheorem{assu}{Assumption}
\newtheorem{prop}{Proposition}
\newtheorem{cro}{Corollary}
\newtheorem*{theorem*}{Theorem}
\newtheorem{example}{Example}
\newtheorem{observation}{Observation}
\newcommand{\exm}{\begin{example}}
\newcommand{\exmm}{\end{example}}
\newcommand{\obs}{\begin{observation}}
\newcommand{\obss}{\end{observation}}
\newcommand{\cor}{\begin{cro}}
\newcommand{\corr}{\end{cro}}
\newtheorem{exa}{Example}
\newcommand{\ex}{\begin{exa}}
\newcommand{\exx}{\end{exa}}
\newtheorem{remak}{Remark}
\newcommand{\rmk}{\begin{remak}}
\newcommand{\rmkk}{\end{remak}}
\newcommand{\thm}{\begin{tm}}
\newcommand{\nt}{\noindent}
\newcommand{\thmm}{\end{tm}}
\newcommand{\lm}{\begin{lma}}
\newcommand{\lmm}{\end{lma}}
\newcommand{\ass}{\begin{assu}}
\newcommand{\asss}{\end{assu}}
\newcommand{\df}{\begin{dfn}  }
\newcommand{\dff}{\end{dfn}}
\newcommand{\prp}{\begin{prop}}
\newcommand{\prpp}{\end{prop}}
\newcommand{\bqu}{\sloppy \small \begin{quote}}
\newcommand{\equ}{\end{quote} \sloppy \large}
\newcommand\cites[1]{\citeauthor{#1}'s\ (\citeyear{#1})}

\newcommand{\eq}{\begin{equation}}
\newcommand{\eqq}{\end{equation}}
\newtheorem{claim}{\it Claim}
\newcommand{\cl}{\begin{claim}}
\newcommand{\cll}{\end{claim}}
\newcommand{\bit}{\begin{itemize}}
\newcommand{\eit}{\end{itemize}}
\newcommand{\ben}{\begin{enumerate}}
\newcommand{\een}{\end{enumerate}}
\newcommand{\bcen}{\begin{center}}
\newcommand{\ecen}{\end{center}}
\newcommand{\fn}{\footnote}
\newcommand{\ds}{\begin{description}}
\newcommand{\dss}{\end{description}}
\newcommand{\prf}{\begin{proof}}
\newcommand{\prff}{\end{proof}}
\newcommand{\cs}{\begin{cases}}
\newcommand{\css}{\end{cases}}
\newcommand{\ml}{\item}
\newcommand{\lb}{\label}
\newcommand{\ra}{\rightarrow}
\newcommand{\tra}{\twoheadrightarrow}
\newcommand*{\supp}{\operatornamewithlimits{sup}\limits}
\newcommand*{\inff}{\operatornamewithlimits{inf}\limits}
\newcommand{\nf}{\normalfont}
\renewcommand{\Re}{\mathbb{R}}
\newcommand*{\mmax}{\operatornamewithlimits{max}\limits}
\newcommand*{\mmin}{\operatornamewithlimits{min}\limits}
\newcommand*{\argmax}{\operatornamewithlimits{arg max}\limits}
\newcommand*{\argmin}{\operatornamewithlimits{arg min}\limits}
\newcommand{\uhr}{\!\! \upharpoonright  \!\! }

\newcommand{\CR}{\mathcal R}
\newcommand{\CC}{\mathcal C}
\newcommand{\CT}{\mathcal T}
\newcommand{\CS}{\mathcal S}
\newcommand{\CM}{\mathcal M}
\newcommand{\CL}{\mathcal L}
\newcommand{\CP}{\mathcal P}
\newcommand{\CN}{\mathcal N}

\newtheorem{innercustomthm}{Theorem}
\newenvironment{customthm}[1]
  {\renewcommand\theinnercustomthm{#1}\innercustomthm}
  {\endinnercustomthm}
\newtheorem{einnercustomthm}{Extended Theorem}
\newenvironment{ecustomthm}[1]
  {\renewcommand\theeinnercustomthm{#1}\einnercustomthm}
  {\endeinnercustomthm}
  
  \newtheorem{innercustomcor}{Corollary}
\newenvironment{customcor}[1]
  {\renewcommand\theinnercustomcor{#1}\innercustomcor}
  {\endinnercustomcor}
\newtheorem{einnercustomcor}{Extended Theorem}
\newenvironment{ecustomcor}[1]
  {\renewcommand\theeinnercustomcor{#1}\einnercustomcor}
  {\endeinnercustomcor}
    \newtheorem{innercustomlm}{Lemma}
\newenvironment{customlm}[1]
  {\renewcommand\theinnercustomlm{#1}\innercustomlm}
  {\endinnercustomlm}

\newcommand{\red}{\textcolor{red}}
\newcommand{\blue}{\textcolor{blue}}
\newcommand{\purple}{\textcolor{purple}}
\newcommand{\mred}[1]{\color{red}{#1}\color{black}}
\newcommand{\mblue}[1]{\color{blue}{#1}\color{black}}
\newcommand{\mpurple}[1]{\color{purple}{#1}\color{black}}

\makeatletter
\newcommand{\customlabel}[2]{%
\protected@write \@auxout {}{\string \newlabel {#1}{{#2}{}}}}
\makeatother


\def\qed{\hfill\vrule height4pt width4pt
depth0pt}
\def\reff #1\par{\noindent\hangindent =\parindent
\hangafter =1 #1\par}
\def\title #1{\begin{center}
{\Large {\bf #1}}
\end{center}}
\def\author #1{\begin{center} {\large #1}
\end{center}}
\def\date #1{\centerline {\large #1}}
\def\place #1{\begin{center}{\large #1}
\end{center}}

\def\date #1{\centerline {\large #1}}
\def\place #1{\begin{center}{\large #1}\end{center}}
\def\intr #1{\stackrel {\circ}{#1}}
\def\R{{\rm I\kern-1.7pt R}}
 \def\N{{\rm I}\hskip-.13em{\rm N}}
 \newcommand{\cprod}{\Pi_{i=1}^\ell}
\let\Large=\large
\let\large=\normalsize


\begin{titlepage}

\def\thefootnote{\fnsymbol{footnote}}
\vspace*{0em}

\title{Binary~Relations~in~Mathematical~Economics:~On~the~Continuity,\vskip .3em Additivity and Monotonicity   Postulates  in  \vskip .5em  Eilenberg, Villegas and DeGroot\fn{This work was initiated during Khan's visit to the Department of Economics, University of Queensland, July 27 to August 13, 2018.    In addition to the hospitality of the Department, Khan also thanks Rabee Tourky for emphasizing the importance of \cite{ei41} during a most pleasant visit  at the trimester program \lq\lq Stochastic Dynamics in Economics and Finance" held by
Hausdorff Research Institute for Mathematics (HIM) in August 2013. The authors are still trying to track down  Lerner (1907).  The authors also thank  Youcef Askoura, Ying Chen, Aniruddha Ghosh and   Eddie Schlee  for conversation and collaboration.  This paper draws its basic conception and composition to an invited plenary talk titled \lq\lq The Role of Positivity in Mathematical Economics: Monotonicity and Free-Disposal in Walrasian Equilibrium Theory", and delivered by Khan at {\it Positivity X} held in Pretoria, July 8-12, 2019. He thanks  Jan Harm and his team of Organizers for the invitation, and for their help in the logistics. He also thanks Jacek Banaciak, Bernard Cornet, Jacobus Grobler, Malcolm,  King, Sonja Mouton, Asghar Ranjbari, Eric Schliesser and Nicoll\`{o} Urbinati for stimulating conversation and encouragement after his talk.     }} 

 \vskip .3em


\author{M. Ali Khan\fn{Department of Economics, Johns Hopkins University, Baltimore, MD 21218.} and  Metin Uyan{\i}k\footnote{School of Economics, University of Queensland, Brisbane, QLD 4072.}}

\vskip 0.5em

\date{July 3, 2020}


\vskip 1.5em

\baselineskip=.18in

\noindent{\bf Abstract:}   This chapter examines  how   {\it positivity} and {\it order} play out in two important questions in mathematical economics, and in so doing, subjects the postulates of {\it continuity, additivity} and {\it monotonicity} to closer scrutiny. Two sets of results are offered:  the first departs from \cites{ei41}    
  necessary and sufficient conditions on the topology under which an anti-symmetric,  complete, transitive and  continuous binary relation exists on a topologically connected space; and the second, from \cites{de70} result concerning an additivity postulate that ensures a complete binary relation on a $\sigma$-algebra to be transitive. These results are framed in the registers of order, topology, algebra and measure-theory; and  also beyond mathematics in  economics: the  exploitation of   Villegas' notion of {\it monotonic continuity} by Arrow-Chichilnisky in the context of Savage's theorem in decision theory, and   the extension of Diamond's impossibility result in social choice theory by Basu-Mitra.  As such,  this chapter has  a synthetic and expository motivation, and  can be read as   a plea for inter-disciplinary conversations, connections and collaboration. 
  
  \hfill  (164~words)

\vskip 2.5em

\noindent {\it 2010 Mathematics Subject} Classification Numbers: 91B55, 37E05.

\vskip 0.6em

\noindent {\it Journal of Economic Literature} Classification
Numbers: C00, D00, D01

\vskip 0.6em


\vskip 0.6em

\noindent {\it Running Title:}  Continuity, Additivity and Monotonicity   Postulates

\end{titlepage}


\tableofcontents

\vspace{15pt}


\vspace{5pt}
\setcounter{footnote}{0}


\setlength{\abovedisplayskip}{0.1cm}
\setlength{\belowdisplayskip}{0.1cm}


\pagebreak

\bqu \textit{It has often happened that a theory designed originally as a tool for the
study of a physical problem came subsequently to have purely mathematical
interest. When that happens the theory is generalized way beyond the point needed for applications, the generalizations make contact with
other theories (frequently in completely unexpected directions), and the
subject becomes established as a new part of pure mathematics. Physics is not the only external source of mathematical theories;
other disciplines (such as economics and biology) can play a similar role.}\fn{\cite[p.419]{ha56}     The part
of pure mathematics so created does not (and need not) pretend to solve
the physical problem from which it arises; it must stand and fall on its own
merits.}  \hfill{Halmos (1956) }   \equ 

\bqu \textit{It is also possible that algebra, as a separate discipline within mathematics may not survive. The 20th century was a period of unification, with algebra invading other areas of math, and they counter-invading it. If I am engaged in studying a family of functions on multi-dimensional manifolds, those families having a group structure, am I working in analysis (the functions), topology (the manifolds) or algebra 
(the groups)?}\fn{\cite[p.319]{de06}}  \hfill{Derbyshire (2006) }   \equ

\section{Introduction} \lb{sec: introduction}

 In this chapter revolving around  the ideas of {\it  positivity} and {\it order} in mathematical economics, one can do worse than begin with  Garett Birkhoff's review of \cite{ei41}: it  is well-worth quoting in full. 

\bqu An  \lq\lq ordered  topological  space"   is,  in  effect,  a  simply  ordered  set  whose  topology  is obtainable by a weakening of its intrinsic topology. The author proves that a topological connected space $X$ can be ordered if and only if the subset of its square $X^2$ obtained by  deleting  the  diagonal  of  points  (x, x)  is  not  connected;  the  same  condition  also characterizes those connected locally connected separable topological spaces which are homeomorphic with subsets of the linear continuum.  \equ

\nt In this,  his paper on \lq\lq ordered topological spaces," \cite{ei41} is justly celebrated for posing  two questions of seminal importance for economic theory.  First, can a continuous binary relation on a set be represented by a continuous function on the same set? Second, what are the conditions on the set under which a complete and continuous relation is necessarily transitive? Both questions, the second perhaps more than the other,  investigate how technical topological conditions, assumed for tractability, necessarily translate into behavioral consequences. However,  
Eilenberg limited himself to the study of anti-symmetric relations, and thereby to studying agency in a context wherein distinct elements in the choice set are necessarily preferred one to another, a kind of extreme decisiveness. It  remained for \cite{de54, de60}  to place  the first question,\fn{In his reproduction of Debreu's theorem [Proposition 1] on the sufficiency of connectedness of a choice set in a finite-dimensional Euclidean space,  \cite{ko72} for example, observes that  \lq\lq Debreu credits a paper of \cite{ei41} as containing the mathematical essence of [his] Proposition 1."}
 and for 
\cite{so65, so67}  the second, in a setting where the symmetric part of  the given binary relation is not an equality, which is to say, the set of  indifferent elements of the relation are not singletons. They and their followers have by now given rise to a rich and mature body of work.  

Eilenberg also asked, and answered,  two other  questions that seem to have had less traction in economic theory, at least in the way that they were initially posed. He asked for conditions on the topology under which there exist \lq\lq nice" relations (in the sense of being anti-symmetric, transitive, complete and  continuous)  on a given set, and furthermore,  turning the matter on its head,  how such relations disallow  sets that are  \lq\lq rich" in the meaning endowed to the term through  the topological and/or algebraic sructures on the set  over which they are defined.  We shall think of these as Eilenberg's third and fourth questions. Both questions are again  natural ones.  The third  is in some sense analogous\fn{Eilenberg's third question is entirely analogous to the existence of a {\it one-to-one} continuous function since he requires the anti-symmetry property. In Section 3 we introduce a result for binary relations that is exactly analogous to the existence of a non-constant continuous function.} to the question concerning conditions on a topology under which non-constant continuous functions exist.  If the topology is too "sparse" then every continuous function is necessarily constant, and every reflexive, transitive and continuous relation is necessarily trivial in sense that no element is preferred to another.   In the context of his fourth question, Eilenberg showed that the existence of a \lq\lq nice" relation  defined on a connected, locally connected and separable space  necessarily renders the space to be a linear continuum.  These results then are a testimony to the mutual imbrication of assumptions on a relation and the space on which the relation  is defined, a two-way relationship that in recent work, Khan-Uyan{\i}k \citeyearpar{ku19a}  see and study as the Eilenberg-Sonnenschein (ES) research program.  

In terms of the third and fourth questions concerning \lq\lq nice" relations, to be sure, topologists have understood this mutual imbrication very well. Thus, for there to be a rich supply of continuous linear functions, the topology on the common domain of the functions must, of necessity, satisfy some properties, and cannot be too sparse. Alternatively, the only continuous functions on a set endowed with an indiscrete topology are the  constant functions;  and digging a little deeper, there is a  plethora of (say) non-locally convex spaces with no continuous function at all other than the zero function.  The question of the existence of a supporting hyperplane is explicitly  studied by \citet{kl63} in the context of an algebraic structure, and in the context of topological vector spaces,    \citet[p. vii]{kpr84} write: 

\bqu The role of the Hahn-Banach theorem may be said
to be that of a universal simplifier whereby infinite-dimensional 
arguments can be reduced to the scalar case by the use of the
ubiquitous linear functional. Thus the problem with non-locally
convex spaces is that of \lq\lq getting off the ground."   \equ 
The point is that there is some {\it hiddenness} in the mutual interaction of a function and set that needs to be flushed out. In terms of the origins,  
\noindent \citet{ur25} studies the problem of determining the most general class of topological spaces in which non-constant real-valued continuous functions exist. \citet{he46} provides an example of a countable, connected Urysohn space\fn{A topological space in which any two distinct points can be separated by closed neighborhoods.} in which every continuous function is constant.  Following Hewitt's work, there are results on the class of topological spaces on which every continuous function is constant; see \citet{ch29} for the original paper, and the following,  for example, for more modern work.  \citet{wa62},  \citet{he65}, \citet{lo95} and   \citet{it86}. 

The question is of substantive consequence  for functional analysis but also beyond it  for economic theory and mathematical economics. 
In terms of this register, the problem gets translated into the question of the sustaining of technologically efficient program as in value maximization programs.  \cite{maj74} furnishes a complete characterization and refers to his result as follows:

\bqu  One should recall that a major motivation behind research
in this area comes from the need to determine whether efficient allocations
can be attained by the use of a price mechanism in a decentralized system
achieving economy of information and utilizing individual incentives.
The implications of any result on complete characterization should be
seriously considered in this context, and as far as [the result] goes, they
seem to be somewhat negative in character. The equivalence established indicates that, in general,
one would need a family of price systems to specify an efficient program.
Indeed, the applicability of the criterion is rather restricted since one has
to know too many prices.\fn{Majumdar continues, \lq\lq  But being a complete characterization,[the result] provides a new angle from which the difficulties faced by
the earlier approaches can be viewed and tends to suggest that simpler
criteria involving fewer price systems, in particular, the use of just one
price system as is typically the case, may be incapable of isolating the set
of efficient programs unless restrictive assumptions on technology are
introduced."   For further work on the problem. see \cite{st86},  and the references to his  chapters  in \cite{fa86}. }  \equ 

It is then to this literature that we connect Eilenberg's third and fourth questions. We see him asking this question: rather than the existence of a a function from a \lq\lq nice" class of functions, does there exist a binary relation from a  \lq\lq nice" class of binary  relations?   And the first contribution of this chapter is that it 
 generalizes Eilenberg's answer to this question by relaxing connectedness and anti-symmetry assumptions:  in a nutshell,  we do  to Eilenberg in this context what Sonnenschein did in another and Debreu did  in yet another. This is to say that we  generalize Eilenberg's result by dropping the anti-symmetry assumption, and then extend the generalization to  $k$-connected spaces, and then to a setting that substitutes  $k$-connectedness with local-connectedness. Finally, we 
  note  that our first two results can be analogously generalized to general preferences, and  connect our results to the literature on the non-existence of non-constant continuous functions.

But we also make another connection that has been missed in the economic literature. This is the application of our results on the   existence of a ``nice" preference relation to  \cites{di65} impossibility theorem:  what  this economic literature sees an an impossibility result,  we see simply as a question of the existence of a nice binary relation  where the adjective nice has been given a meaning and an elaboration in terms of intergenerational equity.  In introducing his own paper, \citet[p. 188]{za07} documents the trajectory of this substantial economic literature.

\bqu
\citet{di65} shows that a complete transitive preference relation that displays intergenerational equity and respects the Pareto ordering cannot be continuous in the topology induced by the supremum norm. \citet{bm03} show that such a preference relation -- whether continuous or not -- cannot be represented by a (real-valued) utility function. On the other hand, \citet{sv80} proves that such preference relations do exist. \cite*{fm03}, \cite*{hetal06}, \cite{bm07}, and \cite*{bss07} provide further results, both positive and negative. 
 \equ 
 Moreover, as already illustrated in Toranzo-Herv{\'e}s-Beleso\citeyearpar{th95}, there are continuous, complete and transitive relations on non-separable spaces which are not representable.  This connection that we make is important in that it sights Eilenberg (1941) as one of the originating papers of this substantial economic literature.  
This concludes our discussion of the first substantive section, Section 3,  of the paper.   
 
Section 4 of the paper returns to Eilenberg's second question: to find a suitable topological condition which ensures the transitivity of a complete, reflexive and continuous binary relations.  \cite{ku19a, ku20a} frame  this question in  settings that remain     squarely   remain within the  purely topological registe, but go considerably beyond   Eilenberg.   In his consideration of the relationship\fn{It is worth noting that both \cite{ei41} and \cite{so65} limit their attention to one way of the two-way relationship, in that they examine the implication  of assumptions on  the choice set on the properties of the relation defined on that set; the {\it backward} direction exploring the implication of the properties of a class of preferences  on the choice set over which they are defined is the signature of the Khan-Uyan{\i}k work.}  however,  \cite{so65}  move to a   a setting that also embrace linear structures.  In a complementary result,    Galaabaatar-Khan-Uyan{\i}k \citeyearpar{gku18}, henceforth GKU,  show the existence of a mixture-continuous,  anti-symmetric, transitive and complete relation defined on a mixture space  renders the setting to be isomorphic to either a  greater-than-or-equal-to relation, or its inverse, defined on the interval!  These results are of substantive consequence for social science since they pertain directly to the formalization of human agency. It mandates that in a sufficiently rich choice set, an agent in an economy, or a player in a game,  cannot be simultaneously consistent (transitive) and extremely decisive (anti-symmetric and complete); or to put the matter in a contra-positive way,  the choice-set of a sufficiently rational agent in the sense of satisfying the above two {\it desiderata}  must of necessity be sparse and impoverished: a linear continuum in the case of Eilenberg and an interval in the case of GKU.  Note that these results, while bearing obvious implication for results on the representation of binary relations, belong to an entirely  different register.  They concern the dove-tailing and mutual imbrication of a set of assumptions on one object for those on a different but not unrelated object.  

The second contribution of this chapter is to make a further move from the register of mixture-spaces to a more abstract algebraic one. Our point of departure now is \cite{vi64, vi67}: this work studied countably additive qualitative probability representations and showed that given a finite additive qualitative probability, monotone continuity  is necessary and sufficient for a countably additive representation\fn{See \citet[Section 5.4.2]{klst71} for further discussion.} It remained for \cite{de70}  to flush out  the abstract algebraic register grounding this result.  The  contribution of Section 4 below  is  (i) to introduce an sharper additivity  postulate, one  supplemented by monotone continuity postulates, on   abstract algebraic structures that are analogous to Villegas' additivity postulate, (ii) to obtain an equivalence result between additivity and transitivity without referring to completeness or continuity of the binary relation, (iii) to show that under additivity, different variants of the monotonicity concepts are equivalent, (iv) to relate our result to Villegas, DeGroot, de Finetti, Arrow and Chichinisky. In particular we highlight the {\it hiddenness} and {\it redundancy}  of the transitivity assumption as these {\it desiderata} are emphasized in \cite{ku19a}.  As such, it  contributes  to the depth and maturity of the ES program.  This concludes our discussion of Section 4 of the paper.

We began this this introduction  by reading \cite{ka41}  as a text revolving around four questions concerning binary relations: leaving Section for notational and conceptual preliminaries,  we shall focus on the third and the fourth in Section 3, and on the second in Section 4. The reader may well wonder our silence about the (first) question that mathematical economists and economic theorists know him by. As mentioned, this pertains to the  representation of a binary relation  by a function, of a continuous relation by a continuous function, of a monotonic relation by a monotonic  function, and of a concave relation by a quasi-concave function. We state the question in all these elaborated way simply to allude the river of work that has accumulated in mathematical economics and mathematical psychology on this question.  But to keep to Eilinberg's parameters except that of his singleton indifference sets, we quote from \cite[p. 3]{be20}. 

\bqu In this expository essay
we consider how much of the theory can be developed from a purely topological
perspective. We focus on those ideas which provide a link between utility theory
and topology, and we leave the economic interpretations to others. Briefly, we give
priority to results that seem to be topologically important, so we pay more attention
to the quotient space of indifference classes than is usual, and more attention to
the order topology than other topologies. \equ

\noindent  We refer the reader to the above article and to  \cite{betal20}, the book of which it is a chapter, and move on.\fn{The reader interested in this (first) Eilenberg question can also see   \cite{wo43}, \cite{na65}, \cite{bm95} \cite{he89b}; we shall return to \cite{wo43} in Section 5.}

\section{Mathematical and Conceptual Preliminaries} \lb{sec: pre}

Let $X$ be a set. A subset $\succcurlyeq$ of $X\times X$ denote a {\it binary relation} on $X.$ We denote an element $(x,y)\in ~\!\!\! \succcurlyeq$ as $x\succcurlyeq y.$ The {\it  asymmetric part} $\succ$ of $\succcurlyeq$ is defined as $x\succ y$ if $x\succcurlyeq y$ and  $y\not\succcurlyeq x$, and its {\it symmetric part} $\sim$ is defined as $x\sim y$ if $x\succcurlyeq y$ and $y\succcurlyeq x.$   
The inverse of $\succcurlyeq$ is defined as  $x\preccurlyeq y$ if $y\succcurlyeq x$. Its asymmetric part $\prec$ is defined analogously and its symmetric part is $\sim$.  We provide the descriptive adjectives pertaining to a relation in a tabular form for the reader's convenience in the table below. 
\vspace{5pt}

\begin{table}[h!]
\begin{center}
\begin{tabular}{lll} 
\hline  \noalign{\vskip 1mm}     
{\it reflexive}   &$x\succcurlyeq x$ $\forall x\in X$\\ 
{\it complete}  & $x\succcurlyeq y$ or $y\succcurlyeq x$ $\forall x,y\in X$\\ 
{\it non-trivial}   &  $\exists x,y\in X$ such that $x\succ y$\\ 
{\it transitive}  & $x\succcurlyeq y\succcurlyeq z \Rightarrow x\succcurlyeq z$ $\forall x,y,z\in X$\\ 
  {\it semi-transitive}  & $x\succ y\sim z \Rightarrow x\succ z$ and  $x\sim y\succ z \Rightarrow x\succ z$ $\forall x,y,z\in X$ \\
  {\it anti-symmetric}  &  $x\succcurlyeq y$ and $y\succcurlyeq x \Rightarrow x=y$ $\forall x,y\in X$ \\
\hline
\end{tabular}
\end{center}
\vspace{-14pt}

\caption{Properties of Binary Relations}
\lb{tbl: relation}
\end{table}  


Let $\succcurlyeq$ be a binary relation  on a set $X$. For any $x\in X$, let    
 $
A_\succcurlyeq(x)=\{y\in X| y\succcurlyeq x\}$ denote the {\it upper section} of $\succcurlyeq$ at $x$  and  $A_\preccurlyeq(x)=\{y\in X| y\preccurlyeq x\}$ its {\it lower section} at $x$.    
Now assume $X$ is endowed with a topology.   
We say $\succcurlyeq$  is {\it continuous} if its upper and lower sections are closed at all $x\in X$ and the upper and lower sections of its asymmetric part $\succ$ are open at all $x\in X$.   
 %


A topological space $X$ is said to be {\it connected} if it is not the union of two non-empty, disjoint open sets. The space $X$ is {\it disconnected} if it is not connected. A subset of $X$ is connected if it is connected as a subspace.    
 We say $X$ is {\it locally connected} if for all $x\in X$, every open neighborhood of $x$ contains a connected and open set containing $x$. 
A {\it component} of a topological space is a maximal connected set in the space; that is, a connected subset which is not properly contained in any connected subset. For any natural number $k,$ a topological space is {\it $k$-connected} if it has at most $k$ components.\fn{See \cite{ku19a} for a detailed discussion on $k$-connectedness.}  
  The concept of  $k$-connectedness provides a quantitative measure of the degree of disconnectedness of a topological space. It is easy to see  that $1$-connectedness is equivalent to connectedness and that any $k$-connected space is $l$-connected for all $l\geq k.$  
\smallskip


\section{On the Existence of a Continuous Binary Relation}

\citet[Theorem I]{ei41} provides a necessary and sufficient condition for the existence of an anti-symmetric,  complete, transitive and continuous binary relation on a connected topological space $X.$ In this section we start with introducing a generalization of Eilenberg's result to $k$-connected spaces,  and then show that when the space is locally connected, then cardinality of the components of the space does not matter. We continue by presenting a result which eliminates the anti-symmetry requirement in Eilenberg's theorem. We end the section with a brief discussion of our results. 

Before  presenting our result, we need the following notation: for any set $X,$ define 
 $$P(X)=\{(x,y)\in X\times X: x\neq y \}.$$ 

\subsection{On Ordered Topological Spaces}

\cite{ei41} calls a topological space {\it ordered} if there exists an anti-symmetric, complete, transitive and continuous binary relation on it. He then presents 

\medskip

\noindent {\bf  Theorem (Eilenberg).} {\it A connected topological space $X$ which contains at least two elements can be ordered if and only if $P(X)$ is disconnected. 
}
\medskip

The following theorem generalizes Eilenberg's theorem to $k$-connected spaces. 

\thm
For any natural number $k,$ a $k$-connected topological space $X$ can be ordered if and only if $P(C)$ is disconnected for each non-singleton component $C$ of $X$. 
\lb{thm: eilorder}
\thmm

\prf[Proof of Theorem \ref{thm: eilorder}] 
 Let $\{C_i\}_{i=1}^\ell$ be the collection of the components of $X$ where $\ell\leq k$.  First note that Theorem (Eilenberg) implies that for each component $C_i$ of $X$ which contains at least two elements, there exists an anti-symmetric,  complete, transitive and continuous binary relation $\succcurlyeq_i$ on $C_i$  if and only if $P(C_i)$ is disconnected.

 In order to prove the forward direction, assume $\succcurlyeq$ is an anti-symmetric,  complete, transitive and continuous binary relation on $X$. Then, for each $C_i,$ the restriction of $\succcurlyeq$ on $C_i,$ defined as $\succcurlyeq_{i}=\succcurlyeq\cap (C_i\times C_i),$ is an anti-symmetric,  complete, transitive and continuous binary relation on $C_i.$ Then,  Theorem (Eilenberg)  implies that $P(C_i)$ is disconnected for all non-singleton $C_i$. 

In order to prove the backward direction, assume $P(C_i)$ is disconnected for each non-singleton $C_i$. It follows from Theorem (Eilenberg) that  there exists an anti-symmetric,  complete, transitive and continuous binary relation $\succcurlyeq_i$ on every non-singleton $C_i$. If $C_i$ is a singleton, then define $\succcurlyeq_i=C_i\times C_i$. Then define a binary relation $\succcurlyeq$ on $X$ as follows: $\bigcup_{i=1}^\ell \succcurlyeq_i ~\!\!\!\! \subseteq ~\!\!\!\!\succcurlyeq$, and for all $i>j$, $C_i\times C_j \subseteq ~\!\!\!\! \succcurlyeq$.  Then, $\succcurlyeq$ is anti-symmetric, complete, and transitive. Since each $C_i$ is closed in $X,$ therefore $\succcurlyeq_i$ has closed sections in both $C_i$ and $X$, and hence $\succcurlyeq$ has closed sections. Therefore, $\succcurlyeq$ is continuous. 
\prff

\thm
A locally connected topological space $X$ can be ordered if and only if $P(C)$ is disconnected for each non-singleton component $C$ of $X$. 
\lb{thm: eilorder2}
\thmm

\prf[Proof of Theorem \ref{thm: eilorder2}] 
Let $\{C_i\}_{i\in I}$ be the collection of the components of $X$.  
%
%
The proof of the forward direction is identical to the proof of the forward direction of Theorem \ref{thm: eilorder}. 
 In order to prove the backward direction, assume $P(C_i)$ is disconnected for each non-singleton $C_i$. It follows from Theorem (Eilenberg) that there exists an anti-symmetric,  complete, transitive and continuous binary relation $\succcurlyeq_i$ on each non-singleton $C_i$. If $C_i$ is a singleton, then define $\succcurlyeq_i=C_i\times C_i$. The well-ordering theorem (\citealp[Theorem, p.65]{mu00}) implies that there exists  an anti-symmetric,  complete and transitive binary relation $\hat \succcurlyeq$ on $I.$ Then define a binary relation $\succcurlyeq$ on $X$ as follows: $\bigcup_{i\in I} \succcurlyeq_i ~\!\!\!\! \subseteq ~\!\!\!\!\succcurlyeq$, and for all $i~\!\hat \succ ~\! j$, $C_i\times C_j \subseteq ~\!\!\!\! \succcurlyeq$. 
 Then, $\succcurlyeq$ is anti-symmetric, complete, and transitive. Since $X$ is locally connected, each $C_i$ is both open and closed. Then, each $\succcurlyeq_i$ has closed sections in both $C_i$ and $X$. Note that for all $C_i$ and all $x\in C_i,$
$$
A_{\succcurlyeq}(x)=A_{\succcurlyeq_i}(x)\cup \left( \bigcup_{j~\! \hat \succ ~\! i} C_j \right)=A_{\succcurlyeq_i}(x)\cup \left( \bigcap_{i~\! \hat \succcurlyeq ~\! j} C_j^c \right).
$$
 Then, it follows from $C_i$ is open for all $i\in I$ that $\succcurlyeq$ has closed upper sections. An analogous argument implies that $\succcurlyeq$ has closed lower sections. Therefore, $\succcurlyeq$ is continuous. 
\prff

\subsection{On Weakly Ordered Topological Spaces}

This subsection provides a necessary and sufficient condition for the existence of a non-trivial, complete, transitive and continuous binary relation on a connected topological space. This result is analogue to Theorem (Eilenberg), except  that the binary relation is not necessarily anti-symmetric. 

\df
A topological space is {\it weakly ordered} if there exists a non-trivial, complete, transitive and continuous binary relation on it.
\lb{df: wordered}
\dff

\thm
A connected topological space $X$ which contains at least two elements can be weakly ordered if and only if $P(X|\sim)$ is disconnected for some equivalence relation $\sim$ on $X$. 
\lb{thm: eilworder}
\thmm

\prf[Proof of Theorem \ref{thm: eilworder}] 
Let $X$ be a topological space with at least two elements. Assume there exists a non-trivial, complete, transitive and continuous binary relation $\succcurlyeq$ on $X$. Let $\sim$ denote the symmetric part of $\succcurlyeq$. Since $X$ is connected, the quotient space $X| \sim$ is connected.  It is easy to show that the induced binary relation $\hat \succcurlyeq$ on $X| \sim$, defined as $([x],[y])\in \hat \succcurlyeq$ if and only if $(x',y')\in ~\!\!\!\succcurlyeq$ for all $x'\in [x]$ and all $y'\in [y]$, is non-trivial, anti-symmetric, complete, transitive and continuous. Then,  it follows from Theorem (Eilenberg) that $P(X| \sim)$ is disconnected. 

In order to prove the backward direction, assume there exists an equivalence relation $\tilde \sim$ on $X$ such that $P(X|\tilde \sim)$ is disconnected. Then, $X|\tilde\sim$ contains at least two elements. 
 Since $X|\tilde\sim$ is connected,  it follows from Theorem (Eilenberg) that there exists an anti-symmetric,  complete, transitive and continuous binary relation $\hat \succcurlyeq$ on $X|\tilde\sim.$ Define a binary relation  $\succcurlyeq$ on $X$ as $(x,y)\in  ~\!\!\!\succcurlyeq$ if and only if $([x],[y])\in \hat \succcurlyeq$. Then the symmetric part $\sim$ of $\succcurlyeq$ is identical to $\hat\sim.$ It follows from $A_{\hat \succcurlyeq}([x])$ and $A_{\hat \preccurlyeq}([x])$ are closed in $X| \tilde\sim$ and the definition of the quotient topology that the sections
 $$
  A_{\succcurlyeq}(x)=\bigcup_{[y] \hat \succcurlyeq([x]} [y] ~\text{ and }~ A_{\preccurlyeq}(x)=\bigcup_{[y] \hat \preccurlyeq [x]} [y]
 $$
of $\succcurlyeq$ are closed in $X,$ hence $\succcurlyeq$ is continuous.  The non-triviality, completeness and transitivity of $\succcurlyeq$ directly follow from its construction . 
\prff
\medskip

Note that in an ordered space, the indifference relation $\sim$ in Theorem \ref{thm: eilworder} is assumed to be the {\it equality} relation.  Hence, as expected, the requirement for the existence of an order is stronger than the requirement for that of a weak order. The following example illustrates a weakly ordered topological space which cannot be ordered. 
\medskip

\noindent {\bf Example.} Let $X=[0,2]$ and the following define a basis for the topology on $X$: $[0,x)$ for all $x\in (1,2],$ $(x,2]$ for all $x\in [1,2),$  and $(x,y)$ for all $x,y\in [1,2].$ Note that the smallest closed set containing any point in $[0,1]$ is $[0,1]$. 
 It is clear that $X$ is connected. Since the topology is not Hausdorff, \citet[1.4]{ei41} implies that there does not exist an anti-symmetric, complete and continuous binary relation on $X$. However, the following is a non-trivial, complete, transitive and continuous binary relation on $X$: $(x,y)\in~\!\!\!\! \preccurlyeq$ for all $x,y\in [0,1]$, and $(x,y)\in ~\!\!\! \preccurlyeq$ for all $x,y\in X$ with $x<y.$

Finally, the methods of proofs presented in Theorems 1 and 2 can be used to provide generalizations of this result to disconnected spaces.

\subsection{Discussion of the Results}

We can apply our results to the literature on the non-existence of a non-constant function on topological spaces as follows.  First, note that every non-constant continuous function induces a non-trivial, complete, transitive and continuous binary relation. Therefore,  by \cites{he46} result we know that there does not exist a non-trivial, complete, transitive and continuous relation. Moreover, note that the space in Hewitt's paper is countable, hence separable, and connected. Therefore, every non-trivial, complete, transitive and continuous relation has a non-constant, continuous real-valued representation.  Therefore, Theorem 3 provides an equivalence condition for the existence of a non-constant function in Hewitt's setting. Hence, Theorem 3 may provide a new perspective on Hewitt's theorem and on the subsequent work in this line of work. 

 Moreover, \cite{mi70}, \cite{go59}, \cite{ki69} and \cite{ja74} provide countable spaces that are connected and satisfy the Hausdorff separation axiom.  Since continuous functions take connected sets to connected sets, therefore there cannot exist a non-constant continuous function on these spaces. We next show that  there does not exist a continuous,  non-trivial, semi-transitive relation with a transitive symmetric part on these spaces. First, by appealing to the current authors' earlier work,\fn{See \citet[Theorem 2]{ku19a}. We refer the reader to  \cite{ku20a} and \cite{uk19b} for generalizations to bi-preference structures and general parametrized topological spaces.} any such relation is complete and transitive. Since the space is countable, it is separable. Therefore, it follows from \citet[Theorem I]{de54} that there exists a continuous real-valued function representing the binary relation. Since the relation is non-trivial, therefore the function is non-constant. This furnishes us a contradiction.

The literature has focused on the existence, or non-existence, of a non-constant continuous function. 
For binary relations, different continuity postulates has been introduced and used in mathematical economics. The existence of  a non-trivial binary relation satisfying different continuity assumptions may be of interest; see  \citet{uk19a} for an extended discussion on the continuity postulate.\fn{There is a literature on different continuity postulates for functions; see \cites{cm16} recent survey on this.}  

\nocite{di65, sv80, bm03, ko72, th95, amt12, la10, za07, de59, ah71, mc02, dw14, kh20} 

\nocite{ch29, ei41, dm41, wa54b, wa54a, wi49, he46, kpr84, se72, mi70, go59, ki69, ja74}

%
%
%
%


\section{On the Additivity Postulate} \lb{sec: ad}

In this section we provide two results on the implications of the additivity postulate. We first show that a strong form of additivity postulate is equivalent to the transitivity postulate. Then we define three monotone continuity postulates on partially ordered sets, inspired by the pioneering work of Villegas  on qualitative probability, and then show that under the additivity postulate, the three continuity postulates are equivalent. We end this section by relating our results to the  antecedent literature. 

\subsection{Additivity and Transitivity: A Two-Way Relationship}


\df
A binary relation $\succcurlyeq$ on an Abelian group $(X,+)$ is called {\it additive} if for  all $x,y,z\in X$, $x\succcurlyeq y$ implies $x+z\succcurlyeq y+z$.  
%
%
%
Moreover, we say $\succcurlyeq$ is {\it strongly additive} if for  all $x_1, x_2, y_1,y_2\in X$, $x_i\succcurlyeq y_i$ for $i=1,2$ implies $x_1+x_2\succcurlyeq y_1+y_2$.  
\lb{dfn: sad}
\dff


We first present a result on the relationship between additivity and strong additivity. 

\prp
Every reflexive and strongly additive relation on an Abelian group is additive. 
\lb{thm: additive}
\prpp

\prf[Proof of Proposition \ref{thm: additive}]
 Assume $\succcurlyeq$ is strongly additive relation on an Abelian group $(X,+)$. Pick $x,y,z\in X$ such that $x\succcurlyeq y$. Then $z\succcurlyeq z$, by reflexivity, and strong additivity of $\succcurlyeq$ imply $x+z\succcurlyeq y+z$. Hence  $\succcurlyeq$ is additive.
\prff
 
 Along with this observation, the next result shows that when a reflexive binary relation is transitive, the two additivity postulates are equivalent. Moreover, it shows that the transitivity of the relation is implied by strong additivity. 

\thm
An additive binary relation $\succcurlyeq$ on an Abelian group $(X,+)$ is transitive if and only if it is strongly additive.
\lb{thm: additivet}
\thmm

\prf[Proof of Theorem \ref{thm: additivet}]
Let $\succcurlyeq$ be an additive binary relation on an Abelian group $(X,+)$. Assume $\succcurlyeq$ is transitive. Pick  $x_1, x_2, y_1,y_2\in X$ such that $x_i\succcurlyeq y_i$ for $i=1,2$. Then it follows from additivity that $x_1+x_2\succcurlyeq y_1+x_2$ and $x_2+y_1\succcurlyeq y_2+y_1$. Then commutativity of  $+$ and transitivity of $\succcurlyeq$ implies that $x_1+x_2\succcurlyeq y_1+y_2$. 

Now assume $\succcurlyeq$ is strongly additive. Pick $x,y,z\in X$ such that $x\succcurlyeq y \succcurlyeq z$.  Then strong additivity implies $x+y\succcurlyeq y+z$. Then additivity  of $\succcurlyeq$ imply $x+y+(-y)\succcurlyeq y+z+(-y)$. Therefore $x\succcurlyeq z$. 
\prff

 The following is a direct corollary of Proposition \ref{thm: additive} and Theorem \ref {thm: additivet}. 

\cor
Every reflexive and strongly additive relation on an Abelian group is transitive.  
\lb{thm: additiver}
\corr

%


\subsection{Implications of Additivity for Monotone Continuity}

Let $(X, \geq)$ be a partially ordered set. We say $X$ is  {\it order-complete} if every non-empty subset of $X$ with an upper bound has a least upper bound. Note that  a poset $X$ is order-complete  if and only if every non-empty subset of $X$ with a lower bound has a greatest lower bound; see Fremlin (3.14B, vol3I).

\df
Let $(X,\geq)$ be an order-complete poset and $\succcurlyeq$ a binary relation on $X$.  We define the following monotone continuity axioms for $\succcurlyeq$. 
\ben[{}, topsep=3pt]
\setlength{\itemsep}{-1pt} 
\ml[{\nf [C1$'$]}]  For all $y\in X$ and all bounded below sequence $\{x_i\}_{i\in \mathbb N}$ in $X$, $x_i\geq x_{i+1} $ and $x_i\succcurlyeq y$ for all $i$ imply  {\nf inf}$\{x_i\}_{i\in \mathbb N}\succcurlyeq y$.
\ml[{\nf [C2$'$]}] For all  $y\in X$ and all bounded above  sequence $\{x_i\}_{i\in \mathbb N}$ in $X$, $x_{i+1}  \geq x_i$ and $y\succcurlyeq x_i$ for all $i$ imply  $y \succcurlyeq \text{\nf sup}\{x_i\}_{i\in \mathbb N}$.
\ml[{\nf [C3$'$]}] For all $y\in X$ and all bounded above  sequence $\{x_i\}_{i\in \mathbb N}$, $x_{i+1}  \geq x_i$ and  $y \prec \text{\nf sup}\{x_i\}_{i\in \mathbb N}$ imply there exists an integer $N>0$ such that, for $i \geq N$, we have $y\prec x_i$. 
\een
\lb{df: mc}
\dff

%
%

%
%
%

\thm 
For any complete and strongly additive binary relation on an Abelian group which is also an order-complete poset, the continuity axioms {\nf C1$'$}, {\nf C2$'$} and {\nf C3$'$} are equivalent. 
\lb{thm: monotone}
\thmm

\prf[Proof of Theorem \ref{thm: monotone}]
Let $(X,+,\geq)$ be an order-complete poset on an Abelian group and $\succcurlyeq$ a complete and strongly additive binary relation on X. It follows from Corollary \ref{thm: additiver} that $\succcurlyeq$ is transitive. 

 First, we show that C1$'$ is equivalent to C2$'$. Note that additivity implies $x\succcurlyeq y$ if and only if $-y\succcurlyeq -x$. Assume C1$'$. Pick  a bounded above sequence $\{x_i\}_{i\in \mathbb N}$ and  $y$ in $X$ such that $x_{i+1}\geq x_{i} $ and $y\succcurlyeq x_i$ for all $i$. Then, $-x_{i}\geq -x_{i+1} $ and $-x_i\succcurlyeq -y$ for all $i$. 
It follows form C1$'$ that {\nf inf}$\{-x_i\}_{i\in \mathbb N}\succcurlyeq -y$.  
We now show that   additivity implies {\nf inf}$\{-x_i\}_{i\in \mathbb N}= -\text{{\nf sup}} \{x_i\}_{i\in \mathbb N}$. Define $\underline x=${\nf inf}$\{-x_i\}_{i\in \mathbb N}$ and $\bar x=-\text{{\nf sup}} \{x_i\}_{i\in \mathbb N}$ Assume towards a contradiction that  $\underline x > \bar x$. By definition, $\underline x\leq -x_i$ for all $i$. Then additivity implies $-\underline x\geq x_i$ for all $i$. Then $-\underline x$ is an upper bound of $\{x_i\}_{i\in \mathbb N}$, hence $-\underline x\geq \bar x$. This contradicts the assumption that $\underline x > \bar x$. An analogous argument yields a contradiction for $\underline x < \bar x$. Therefore, $\underline x = \bar x$.     
Then $-\text{{\nf sup}} \{x_i\}_{i\in \mathbb N} \succcurlyeq -y$, hence by additivity,  $y\succcurlyeq \text{{\nf sup}} \{x_i\}_{i\in \mathbb N}$. Therefore,  C2$'$ holds. The proof of the converse relationship is analogous.

We next show that  C2$'$ is equivalent to C3$'$. Assume  C2$'$. Assume towards a contradiction that there exists a  bounded above $\{x_i\}_{i\in \mathbb N}$ and $y$ in $X$ such that $x_{i+1}  \geq x_i$ and  $y \prec \text{\nf sup}\{x_i\}_{i\in \mathbb N}$, but for all $N>0$, there exists $j\geq N$ such that $y\succcurlyeq x_i$. Then there exists a subsequence  $\{x_i{_k}\}_{k\in \mathbb N}$ such that for all $k$, $x_{i_{k+1}}  \geq x_{i_k}$  and $y\succcurlyeq x_{i_k}$. Then C2$'$ implies  $y \succcurlyeq \text{\nf sup}\{x_{i_k}\}_{k\in \mathbb N}$.  It is easy to see that $\text{\nf sup}\{x_{i_k}\}_{k\in \mathbb N}=\text{\nf sup}\{x_{i}\}_{i\in \mathbb N}$. This contradicts the assumption that  $y \prec \text{\nf sup}\{x_i\}_{i\in \mathbb N}$. Hence C3$'$ holds. The converse relationship immediately follows from the definitions. 
\prff

\subsection{Discussion of the Results}

\cite{vi64} introduced the following additivity concept  for binary relations on a $\sigma$-algebra. 

\df
A preference relation $\succcurlyeq$ on a $\sigma$-algebra $\mathcal X$ on a set $X$ is {\it Villegas-additive} if for all $A_1, A_2, B_1, B_2\in \mathcal X$ with $A_1\cap A_2=B_1\cap B_2=\emptyset$,  $A_i\succcurlyeq B_i$ for $i=1,2$ implies $A_1\cup A_2 \succcurlyeq B_1\cup B_2$. If, in addition, $A_1\succ B_1$ or $A_2\succ B_2$, then $A_1\cup A_2 \succ  B_1\cup B_2$. 
\dff

\smallskip

\nt First note that the union operation is similar to the additivity operation\fn{Note that \citet[p. 336]{fi86} calls Villegas-additivity the {\it additivity axiom}.}   but it does not satisfy all properties the addition in an Abelian group satisfies. Moreover, the usual additivity assumption is neither stronger nor weaker than Villegas-additivity: the latter imposes restriction on a smaller class of elements whereas additivity does not impose a restriction on the strict relation.   \citet[Theorem 1, p. 71]{de70} followed Villegas and proved a result  analogous to Theorem \ref{thm: additivet} where the space is a $\sigma$-algebra with the usual inclusion relation. 

\medskip

\nt  {\bf Theorem} (DeGroot).   {\it Every complete and Villegas-additive binary relation on a $\sigma$-algebra is transitive.}

\medskip


We next apply our results to de Finetti's expected utility representation theorem.  Let $X=\Re^n$ 
 which is endowed with the usual topological, algebraic and order structures. 
%
   A real valued function $u$ is called {\it monotone} if for all $x,y\in \Re^n$ such that $x> y$ (i.e. $x_i\geq y_i$ for all $i$ and $x\neq y$), $u(x) > u(y.$) 
   A preference relation $\succcurlyeq$ on $\Re^n$ is {\it monotone} if for all $x,y\in \Re^n,$ if $x>y,$ then  $x\succ y.$
The following theorem is due to \citet{de37, de74}.\fn{See  \citet[Theorem A.2.1, p.161]{wa89} for the statement and further details.}

\medskip

\nt  {\bf Theorem} (de Finetti). {\it 
 Let $\succcurlyeq$ be a binary relation on $\Re^n.$ The following are equivalent. 
\ben[{\nf (a)},  topsep=-1pt]
\setlength{\itemsep}{-1pt} 
\ml The binary relation $\succcurlyeq$ is complete, transitive, additive and continuous. \lb{it: dfp}
\ml There exist positive $(p_i)_{i=1}^n,$ summing to one, such that $u(x)=\sum_i p_i x_i$ represents $\succcurlyeq.$ \lb{it: dfu}
\een
}

\medskip
The equivalence theorem of de Finetti can be restated as 
\cor
 Let $\succcurlyeq$ be a binary relation on $\Re^n.$ The following are equivalent. 
\ben[{\nf (a)},  topsep=-1pt]
\setlength{\itemsep}{-1pt} 
\ml The binary relation $\succcurlyeq$ is complete, strongly additive and continuous. 
\ml There exist positive $(p_i)_{i=1}^n,$ summing to one, such that $u(x)=\sum_i p_i x_i$ represents $\succcurlyeq.$
\een
\lb{thm: definettic}
\corr

\medskip

\nt Therefore, we can drop the transitivity assumption in de Finetti's theorem by replacing additivity with strong additivity, which are equivalent in the presence of the transitivity postulate. We can also drop the completeness assumption; see \cite{uk19a} for a detailed exposition on the hiddenness and redundancy in mathematical economics.\fn{See also  \citet[Section 5.4.2]{klst71} for an interesting discussion on  hiddenness and redundancy. Moreover, it may be of interest to generalize this result to groupoids or semigroups; see \citet[Chapter 11]{fi72}.} 



We next move to monotone continuity.  The second subsection above is an attempt to understand  this postulate introduced in Villegas, DeGroot, Arrow and Chichilnisky  in order  to study  qualitative/subjective probability. As we illustrate above, monotone continuity neither requires any topological property on the choice set, nor uses the structure of the unit interval, unlike the continuity assumption of \cite{hm53}. Hence, an investigation of the relationship between monotone continuity with the other continuity postulates, and its applications may be of interest, and our definitions and results can be considered as the first step for such investigation.\fn{See \cite{ku19a} and \cite{gku20b} for a discussion on the relationship among different continuity postulates. Moreover,   DeGroot's assumption SP$_5$ ``There exists a random variable which has a uniform distribution on the interval $[0,1]$"  may have some relevance to the existence of a nice relation, or a continuous function. We leave this for future work.}    \cite{vi64} and \cite{de70} provide the following monotone continuity postulates for binary relations defined on $\sigma$-algebras. 

\df
Let $\mathcal X$ be a $\sigma$-algebra on a set and $\succcurlyeq$ a binary relation on $\mathcal X$. We define the following monotone continuity axioms for $\succcurlyeq$. 
\ben[{}, topsep=3pt]
\setlength{\itemsep}{-1pt} 
\ml[{\nf [C1]}] For all sets $\{A_i\}_{i\in \mathbb N}, B$ in $\mathcal X$,   $A_1 \supseteq A_2\supseteq \cdots$ and $A_i\succcurlyeq B$ for all $i$ imply $\bigcap_i A_i\succcurlyeq B$.
\ml[{\nf [C2]}] For all sets $\{A_i\}_{i\in \mathbb N}, B$ in $\mathcal X$,   $`A_1 \subseteq A_2\subseteq \cdots$ and $B\succcurlyeq A_i$ for all $i$ imply $B\succcurlyeq \bigcup_i A_i$.
\ml[{\nf [C3]}] For all sets $\{A_i\}_{i\in \mathbb N}, B$ in $\mathcal X$,   $A_1 \subseteq A_2\subseteq \cdots$ and $B\prec \bigcup_i A_i$ imply there exists an integer $N>0$ such that, for $i \geq N$, we have $B\prec A_i$. 
\een
\dff

The following is a result analogous to Theorem \ref{thm: monotone} above for the special case of $\sigma$-algebras. 

\thm
For any complete and Villegas-additive binary relation on  a $\sigma$-algebra, the monotone continuity postulates {\nf C1}, {\nf C2} and {\nf C3}  are equivalent. 
\lb{thm: mcequiv}
\thmm

\nt  The equivalence between {\nf C2} and C3 is due to \citet[Theorem]{vi64} and between  C1 and C2 is due to \citet[Theorem 5]{de70}. 

\smallskip

\cite{vi64, vi67} studied countably additive qualitative probability representation and showed that given a finitely additive qualitative probability, monotone continuity  is necessary and sufficient for countably additive representation;  see \citet[Section 5.4.2]{klst71} for further discussion. In particular, the following result is quoted.\fn{For definitions, we refer the reader to \citet{klst71}.}

 \medskip
 
\nt {\bf Theorem (Villegas).}  {\it A finitely additive probability representation of a structure of qualitative probability, on a $\sigma$-algebra, is countably additive if and only if the structure is monotonically continuous. }

\bigskip

Finally, the following monotone continuity postulate is due to \cite{ar71}. 

\df
Given $a$ and $b$, where $a\succ b$, a consquence $c$ and a vanishing sequence $\{E_i\}$, suppose sequence of actions satisfy the conditions that $(a^i,s)$ yield the same consequences as $(a,s)$ for all $s\in E_i^c$, and the consequence $c$ for all $s\in E_i$, while $(b^i,s)$ yield the same consequences as $(b,s)$ for all $s\in E_i^c$, and the consequence $c$ for all $s\in E_i$. Then, for all $i$ sufficiently large, $a^i\succ b$ and $a\succ b^i$. 
\dff

%

\nt \citet{ch10} interpreted Arrow's definition as follows and showed  that it is equivalent to the continuity postulate C1. 

\df
Let $\mathcal X$ be a $\sigma$-algebra on a set and $\succcurlyeq$ a binary relation on $X$. 
 We call $\succcurlyeq$ satisfies {\it Monotone Continuity Axiom 4} (C4) if for all  $\{A_i\}_{i\in \mathbb N}, F, G$ in $\mathcal X$,   $A_1 \supseteq A_2\supseteq \cdots$, $\bigcap_{i=1}^\infty A_i=\emptyset$ and $F\succ G$  imply there exists $N>0$ such that altering arbitrarily the events $F$ and $G$ on the set $A_i$, where $i > N$, does not alter the  ranking of the events, namely $F'\succ G'$, where $F'$ and $G'$ are the altered events.
\dff

\smallskip

%


\section{Order and Positivity in Mathematical Economics }
We began this essay with Halmos' take on how applied mathematics transits to pure mathematics; and Derbyshire's take on how an important  sub-field, with increasing importance, gets incorporated into the larger field of which it is a part, and thereby changes the identity of the larger field and loses its own. In this concluding section\fn{In his participation in the composition of this section, Khan should like to acknowledge his indebtedness  to conversations with Malcolm King, and Niccol\`{o} Urbinati,  and to JJ Grobler's inspiring talk titled {\it  101 years of vector lattice theory:
A general form of integral: PJ Daniell (1918)} at  the Conference. He should also like to acknowledge the stimulus received from Schliesser's readings of \cite{fo08}. }   to this chapter on binary relations in mathematical economics, we read, against the grain,  these two texts and their claims      on the incorporation of {\it positivity} and {\it order-theoretic methods} in  Walrasian general equilibrium theory.

In  classical Walrasian general equilibrium theory,  as brought to fruition in  \cite{ko57}, \cite{de59}, \cite{ni68}, \cite{ah71}  and \cite{mc02}, the agents in the economy are categorized as {\it consumers} and {\it producers,}  with the former
parametrized by preferences (a binary relation) defined on a (consumption) set and endowments being elements of such a set; and the latter  drawing their signature simply by having an  access to a {\it production set.}\fn{Our choice of these four texts should perhaps be justified.  For the texts of Debreu and McKenzie, we can appeal to \cite{dw14} who argue that the 1954 papers of Arrow-Debreu and McKenzie wrought fundamental changes in economic theory, a claim contested in   \cite{kh20} who urged the inclusion of Uzawa, Nikaido and Gale also as  fellow-pioneers of what we are calling here \lq\lq Walrasian general equilibrium theory." The naming of the Arrow-Debreu model or the Arrow-Debreu-Mckenzie model facilitated a homogeneous monolithic view and added to the cofusion and to an unfortnate haste in canonization; see Footnote~\ref{fn:texts1} below.  A confounding factor in this is that many of the pioneers of Walrasian general equilibrium theory were also pioneers of linear and non-linear programming; Uzawa breing one of the leaders. For this line of work, see \cite{dss58}, and the recent application of Uzawa's consequential extension of the Kuhn-Tucker-Karush theorem in \cite{kss19}. \label{fn:texts} }    The vernacular of {\it order} and {\it positivity}   is relevant in so far as it is relevant to its  constituent conceptions of a consumer and a producer. The idea of {\it monotonicity} enters the theory of production  through the assumption of {\it free disposal,} an assumption delineated by   \cite{de54} in the context of a production set, say  in a ordered normed space whose positive cone has a non-empty interior.  

\bqu The assumption of free disposal for the technology means that if an input-output combination is possible, so is one where one where some outputs are smaller or some inputs larger; it is implied that a surplus can be freely disposed of. With this assumption, if the production set is non-empty, it has an interior point.  \equ   

\nt It is the existence of an interior that proves crucial for the sustainablity of technologically efficent production plans and Pareto optimal allocations  through individual value and profit maximization.\fn{We invite the reader to compare Debreu's definition with corresponding definitions of the concept in the five texts  to which Footnote~\ref{fn:texts} refers. The idea of ``free disposal" is intimately tied to the non-negativity of prices; see \cite{ha05}, and compare \cite{de59} and \cite{mc02} on this issue.}

As far as the theory of the consumer is concerned, the ideas of {\it order} and {\it positivity} enter through the assumption of {\it monotonicity} of preferences  which gets translated into \lq\lq more is always preferred to less." To be sure,  it factors into the Eilenberg questions regarding binary relations with which we began  the introduction. Thus    \cite[p. 106]{ah71} write: 

\bqu Wold seems to have been the first to see the need of specifying assumptions under which the representation of the continuous utility functions exists.  Wold assumed that the [consumption set] is the entire non-negative orthant [of  finite-dimensonal Euclidean space] and that preference is strictly monotone in each commodity.  A very considerable generalization, based on a mathematical paper by \cite{ei41}, was  achieved with  the deeper methods of  \cite{de54}; he assumed only the continuity of preferences  and the connectedness of the [consumption set] (a property weaker than convexity).   \equ 

\nt This is an important passage:  its irony lies in the fact that it comes from two of the more distinguished and senior Walrasian theorists at the time who could not refrain from drawing arbitrary and needless distinctions between   mathematicians  and economists, and between mathematical and economic papers, and thereby in sighting \cite{ei41}, and devaluing it at the same time. The point is that Eilenberg and Wold  were independent pioneers of what later assumed the identity of an important subfield of  \lq\lq choice and decision theory."\fn{We can recommend \cite{fi72}   \cite{lu00}   \cite{gi04,gi09}. \cite{mo19} and their references for this subject, which  branches off also into mathematical psychology. \label{fn:dtexts} }  

But returning to trajectories being implicitly charted by  Halmos and to Derbyshire, the point is that the monotonicity assumption for consumers in the Walrasian conception comes rather late in its  development: it is not there, for example,  in \cite{de59}, or in \cite{mc02}, or the term even  indexed in \cite{ah71}.\fn{This also suggests how much a reader of Walrasian general equilibrium theory loses by ascribing to it a monolithic conception. Each of these pioneers had their own ways of looking at their subject. In this connection one may also refer the interested reader to McKenzie's conception of production in his \cite[Section 2.8, pages 77-82]{mc02} on an  \lq\lq Economy of Activities."  It is also perhaps worth noting that Debreu's resistance to the monotonicty assumption on  consumers   may be due to his having relaxed the montonicity assumption in \cite{wo43}. To the authors knowledge, his first recourse to the assumption is in connection with the Debreu-Scarf theorem in 1963, and to be sure the auumption irrevocably enters into the field with Aumann and his Israeli school of Walrasian theory; see \cite{de83} for the relevant papers and references.  As emphasized in \cite{kh20}, the erasure of the production sector can also be ascribed to this school, and it becomes folded into the iideological divide between the \lq\lq two Cambridges," those of the UK and the US. \cites{fo08}  emphasis on \lq\lq governmentality" in the formulation of {\it perfect competition} and its normative properties is clearly relevant here.  \label{fn:texts1}} The more important question, however is where the subject is in terms of these, their trajectories. This is a question that merits an investigation of its own, and is outside the scope of this technical essay: it suffices to make two observations.  With respect to Halmos, classical Walrasian general equilibrium theory has neglected, by its very definitional conception, interdependencies between the parametrizations of what it sees as the relevant agents in the economy;  and classical game theory, again by virtue of its definitional conception, has neglected the market in its formalizations.   The applied problems of our time cry out for a  formalization of these interdependencies  in what perhaps ought to be a synthetic view of both subjects. Thus even after 70 years, mathematical economics (including game theory) has very much retained its dependence on both economics {\it and}  mathematics. This is to say that it has remained pure {\it and} applied. 
As to Derbyshire on algebra,  in terms of the algebraic approach to these subjects, it has yet to be incorporated into both Walrasian general equilibrium theory  and in non-cooperative game theory. In the authors' judgement, this cannot but be a fruitful task.

There is another, perhaps narrower,  way to view the substance of these results.   The question of the \lq\lq right" commodity space for general Walrasian general equilibrium, or the \lq\lq right setting" of the individual action sets in  game theory,  has not been explicitly posed. There has been little need to do so.  Given the substantive questions at issue, the economic or game-theoretic formulations assume a strong-enough structure on the payoff functions and the choice sets by setting them either in a finite-dimensional Euclidean space, or in the context of game theory, a finite number of actions,  to allow the question to be investigated and determinatively answered. When this rather arbitrary limitaion is removed, the question becomes of consequence, and notions of {\it order} and {\it positivity} began to take on colours that one may not have previously  imagined.\fn{This investigation remains an ongoing   project of Khan and Urbinati, and in his talk in Pretoria, Khan made some room to expand at some length to report on Nikaido's contributions to this question in keeping with this project.}

This   introduction   has framed the results to follow as stemming from \cites{ei41} seminal work. In this connection we observe that it is an interesting curiosum in the history of ideas that a piece of work entirely peripheral to an author's {\it oeuvre,} written almost as a fragmentary passing thought,  proves to be of such decisive and sustainable consequence in what may have been  perceived at the time of its writing to be an unrelated discipline. Eilenberg's paper, along with \cites{ka41} fixed point theorem,  coincidentally published in the same year, may well be two canonical examples.\fn{1941 was a particular productive years for Eilenberg, especially given the standards of the time: in addition to the paper being discussed here, he published at least six other papers; see \cite[p.302]{de06} for his 1940 meeting with Saunders MacLane, and his subsequent involvement with algebraic topology. His paper with Wilder on \lq\lq uniform local connectedness and contractability" was to follow an year later, and the generalization of Kakutani's fixed point theorem with Montgomery, five years later.}  In any case, as far as mathematical economics is concerned, the belated recognition of this pioneering paper  by \cite{de54, de64} and \cite{so65, so67} has subsequently waned, and it is only recent work that has re-emphasized Eilenberg's work and given it  importance under the rubric of what it refers to as the Eilenberg-Sonnenschein program.   It is a source of satisfaction to the authors that its importance  can also be delineated in a chapter on a book on  {\it positivity} and {\it order}.


\nocite{ei41, wo43, bi48, de54, na65, he89a, he89b, he91, bm94, bh19, betal20, bz20, be20}

\bigskip



%
 
\bigskip

\setlength{\bibsep}{5pt}
\setstretch{1.1}


\nocite{ar65, ar66, ar72, de70, vi64, vi67, ch10, hm53, uk19a, ku20a, gku19}


\bibliographystyle{../../References/econometrica} 
\bibliography{../../References/References.bib}

\end{document}